\documentclass[manuscript]{acmart}

\AtBeginDocument{%
  }

\usepackage{xspace}
\usepackage{booktabs}
\usepackage{dirtytalk}
\usepackage{cleveref}
\usepackage{multirow}
\usepackage[inline]{enumitem}
\usepackage{ifthen}
\usepackage{xcolor}
\usepackage{subcaption}

\title{\system: Writing with Crowds Alongside AI}

\newcommand{\crowd}{Crowd\xspace}
\newcommand{\plot}{Story Plot Generator\xspace}
\newcommand{\gptplot}{GPT3-Plot\xspace}
\newcommand{\gptc}{GPT3-Continuation\xspace}

\newcommand{\system}{\textsc{Inspo}\xspace}
\newcommand{\oldSystem}{\textsc{Heteroglossia}\xspace}

\newcommand{\soc}{sociocultural\xspace}
\newcommand{\Soc}{Sociocultural\xspace}

\definecolor{ao(english)}{rgb}{0.0, 0.5, 0.0}

\newcommand{\avon}[1]{{\small\textcolor{red}{\bf [#1 --Avon]}}}
\newcommand{\kenneth}[1]{{\small\textcolor{blue}{\bf [#1 --Kenneth]}}}

\newcommand{\eg}{{\it e.g.}}
\newcommand{\ie}{{\it i.e.}}

\newenvironment{myquotewithindent}%
  {\list{}{\leftmargin=0.1in\rightmargin=0.1in}\item[]}%
  {\endlist}

\newcommand{\participant}[1]{%
    \ifnum#1=1 P001\xspace \else
    \ifnum#1=2 P002\xspace \else
    \ifnum#1=4 P003\xspace \else
    \ifnum#1=5 P004\xspace \else
    \ifnum#1=10 P005\xspace \else
    \ifnum#1=11 P006\xspace \else
    \ifnum#1=14 P007\xspace \else
    \ifnum#1=15 P008\xspace \else
    Unknown\xspace
    \fi\fi\fi\fi\fi\fi\fi\fi
}

\begin{document}


\author{Chieh-Yang Huang}
\affiliation{%
  \institution{MetaMetrics Inc.}
  \city{Durham}
  \state{NC}
  \country{USA}
}
\email{cyhuang@lexile.com}

\author{Sanjana Gautam}
\affiliation{%
  \institution{University of Texas at Austin}
  \city{Austin}
  \state{TX}
  \country{USA}
}
\email{sanjana.gautam@utexas.edu}

\author{Shannon McClellan Brooks}
\affiliation{%
  \institution{Pennsylvania State University}
  \city{University Park}
  \state{PA}
  \country{USA}
}
\email{sum1423@psu.edu}

\author{Ya-Fang Lin}
\affiliation{%
  \institution{Pennsylvania State University}
  \city{University Park}
  \state{PA}
  \country{USA}
}
\email{yml5563@psu.edu}

\author{Tiffany Knearem}
\affiliation{%
  \institution{Google}
  \city{Cambridge}
  \state{MA}
  \country{USA}
}
\email{tknearem@google.com}

\author{Ting-Hao `Kenneth' Huang}
\affiliation{%
  \institution{Pennsylvania State University}
  \city{University Park}
  \state{PA}
  \country{USA}
}
\email{txh710@psu.edu}

\renewcommand{\shortauthors}{Huang et al.}

\begin{abstract}
    The use of artificial intelligence (AI) to support creative writing has bloomed in recent years. 
However, it is less well understood how AI compares to on-demand human support.
We explored how writers interact with both AI and crowd worker writing assistants in creative writing. 
We replicated the interface of the prior crowd-writing system, \oldSystem, and developed \system, a text editor allowing users to request suggestions from AI models and crowd workers. 
In a one-week deployment study involving eight creative writers, we examined how often participants selected crowd workers when fluent AI text generators were also available. 
Findings showed a consistent decline in crowd worker usage, with participants favoring AI due to its faster responses and more consistent quality. 
We conclude with suggestions for future systems, recommending designs that account for the unique strengths and weaknesses of human versus AI assistants, strategies to address automation bias, and \soc views of writing.

\end{abstract}

\received{20 February 2007}
\received[revised]{12 March 2009}
\received[accepted]{5 June 2009}

\maketitle


\section{Introduction and Backgrounds}

Large language models (LLMs) are currently at the forefront of writing assistance research in HCI~\cite{lee2024design}.
Their ability to generate fluent text and follow user instructions has sparked numerous projects in recent years, exploring how these capabilities might impact or support users' writing. 
Looking at the broader context of HCI, this is not the first wave of research focused on leveraging technology to assist writing. 
Before computers could fluently generate text in response to open-ended prompts, researchers were already pushing the boundaries of what writing assistants could achieve~\cite{greer2016introduction}.
A notable example is the development of ``crowd writing'' systems~\cite{feldman2021we}.
These systems utilized online crowd workers, recruited programmatically from microtask platforms like Amazon Mechanical Turk, to perform tasks similar to those now handled by LLMs: 
producing text based on provided information and instructions.
Such crowd-writing systems allowed researchers to study how users interacted with writing assistants that provided capabilities beyond the automated technology available at the time.
For instance,
\textsc{Soylent} explored how users can write using an editor that exhibited human-like intelligence~\cite{bernstein2010soylent};
\textsc{Knowledge Accelerator} divided complex writing tasks into smaller pieces, which could be completed individually, and then reassembled into a cohesive final product~\cite{kittur2011crowdforge,hahn2016knowledge};
\oldSystem explored how writers can work with story ideas generated from different characters' perspectives~\cite{heteroglossia-huang-2020}.
Like many crowd-powered systems, crowd-writing systems often share the vision of the ``human computation'' paradigm, as characterized by~\citeauthor{quinn2011human}: 
the functions operated by human workers would be eventually ``solvable by computers.''
The use of crowd workers (temporarily) bridged the socio-technical gap between users' needs and system capabilities~\cite{ackerman2000intellectual}, enabling researchers to study user interaction and design challenges in anticipation of future technological advancements.
The emergence of capable LLMs centers the question of  
\textit{should we stop building crowd-writing systems?}

In this paper, we focus on creative writing, particularly story writing. 
Our key stance is that \textbf{humans generally are still much better at creative writing than the best LLMs.} 
Studies have shown that human-written stories are more creative~\cite{bellemare2024divergent,tian2024large} and generally of higher quality~\cite{10.1145/3613904.3642731}, 
LLMs struggle with comedy writing~\cite{mirowski2024robot}, and 
humans are also more effective at editing story drafts~\cite{chakrabarty2024can}.
Therefore, in theory, crowd workers should still be capable of providing higher-quality support for story writers than LLMs, as humans are generally better at writing and editing stories.
However, the reality is more complex. 
When considering literature comparing stories written particularly by online crowd workers against that of LLMs, findings are mixed: 
some studies suggest that crowdsourced stories offer more imaginative scenarios, settings, and rhetoric~\cite{begus2023experimental},
while others show no differences~\cite{orwig2024language}.
Furthermore, despite LLMs generally not matching human ability in creative writing, certain steps of the writing process, such as overcoming writer's block or maintaining consistency~\cite{kreminski-martens-2022-unmet}, have shown to benefit from computational assistance~\cite{10.1145/3635636.3656201}. 
It would be unwise to exclude AI from writing systems simply because it is not yet on par with human capabilities.
As LLMs continue to develop, how crowd-writing systems can evolve to integrate computational support while preserving the advantages of human involvement is becoming an increasingly important research area.

\begin{figure*}[t]
    \centering
    \includegraphics[width=0.8\linewidth]{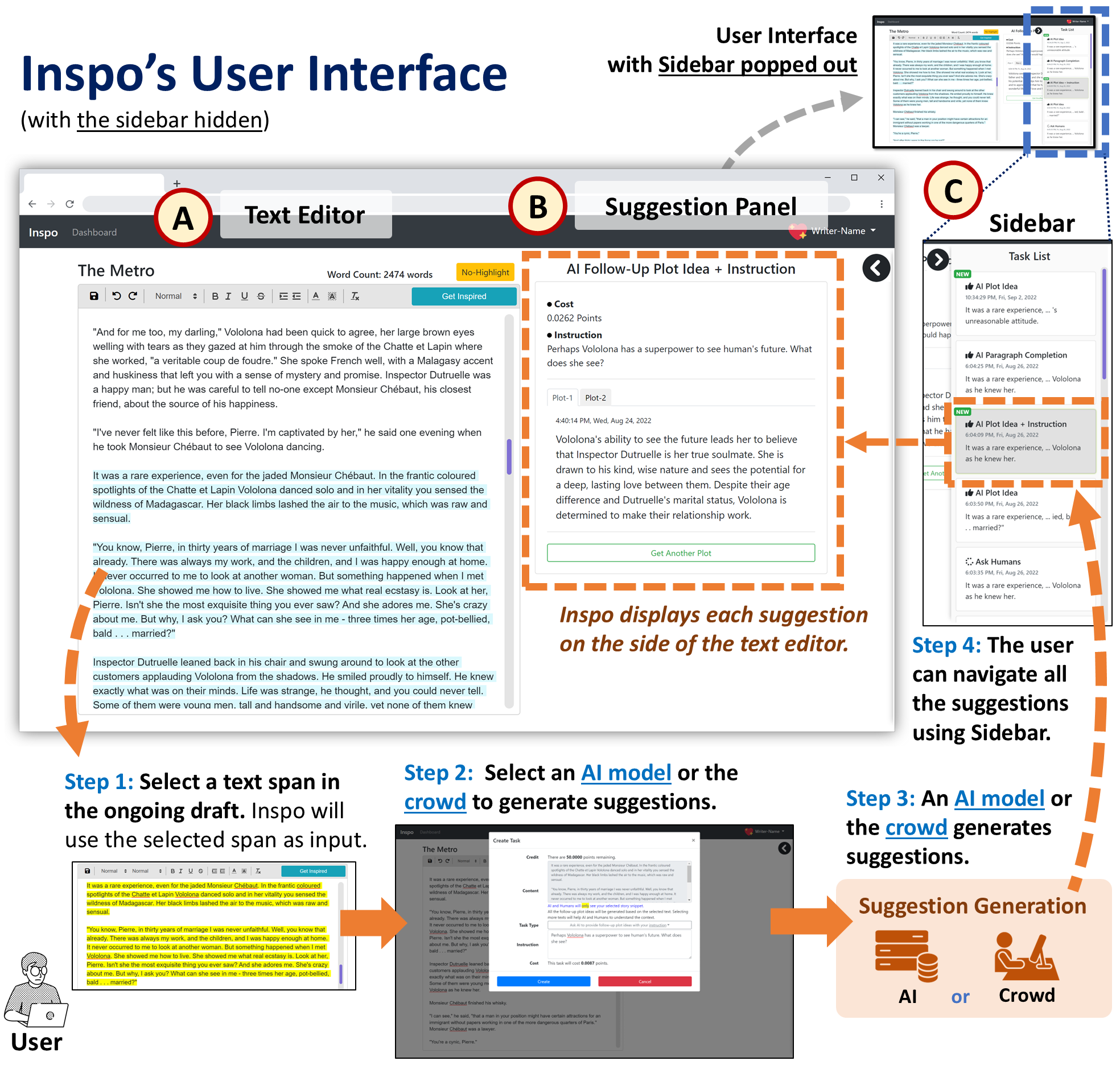}
    \vspace{-4mm}
    \caption{\system contains (A) a text editor, (B) a suggestion panel, and (C) a sidebar. Users can select any part of a working draft and send it to \system to request suggestions from either the text-generation model of their choice (including GPT-3) or online crowd workers.}
    \label{fig:interface}
    \vspace{-4mm}
\end{figure*}

As a step toward understanding AI's impact on crowd-writing systems, 
this paper examines writers' choices and interactions when presented with the option of using either crowd workers or AI as writing assistants.
Specifically, we replicated the interface and in-situ suggestion delivery design of a prior crowd writing system, \oldSystem~\cite{heteroglossia-huang-2020}, and developed \textbf{\system}, a 
text editor that allows 
writers to request suggestions from both AI models and online crowd workers (Figure~\ref{fig:interface}).
Writers using \system can select a portion of their draft and request suggestions from four different sources:
\textbf{{\em (i)} crowd workers (\crowd), 
{\em (ii)} a story plot generation model (\plot), 
{\em (iii)} GPT-3 for generating follow-up story arcs (\gptplot), and 
{\em (iv)} GPT-3 for continuing the given story (\gptc)}.
Through a one-week deployment study involving eight creative writers, 
we explored how often participants selected the crowd option when fluent text generators were also available \textbf{(RQ1)}. 
Additionally, by analyzing their writing logs and user behavior data within \system, we examined how the suggestions generated by crowds and AIs affected users' writing quantity \textbf{(RQ2)} and content \textbf{(RQ3)}, respectively.

Our findings revealed that, over the course of the study, participants increasingly favored AI over crowd workers. 
Writing log analysis showed that \gptc had the greatest impact on both the word count and content of the stories. 
Interviews further indicated that participants stopped using the crowd option due to slower response times and inconsistent quality, despite some unique benefits like the ``human touch'' in crowd responses. 
These results suggest that future crowd-writing systems---or crowd-AI hybrid systems---need to account for the distinct strengths of human and AI support.
In the case of \system, both crowd and AI responses were delivered through the same interface, but the immediacy of AI made it more appealing to users. 
To conclude, we offer design suggestions advocating for systems that intentionally differentiate between human and AI assistants, optimizing the strengths of both.

\section{\system System}



Figure~\ref{fig:interface} shows \system's user workflow and interface, which follow the design of \oldSystem by displaying suggestions beside the text editing panel and using highlighted colors to associate suggestions with the corresponding selected input text spans.
The text editor (\Cref{fig:interface}A) allows users to write stories and request suggestions by selecting a text snippet and clicking the ``Get Inspired'' button. 
This opens the Task Creation Panel (Step 2 in \Cref{fig:interface}), where users can choose a generation method and set task-specific parameters like additional instructions or the number of suggestions. 
Upon clicking the ``Create'' button, \system generates suggestions and displays them in the Sidebar and Suggestion Panel.
The Sidebar (\Cref{fig:interface}C) lists all suggestion tasks for the document, showing the task type and a brief snippet of the corresponding draft, and clicking on a task brings up the suggestion's content in the Suggestion Panel and links it to the relevant text snippet.
The Suggestion Panel (\Cref{fig:interface}B) shows the generated suggestions and allows users to request more with the same settings by clicking the ``Get Another Plot'' button. 
Each generated outcome is displayed in separate tabs (\eg, ``Plot-1'' and ``Plot-2''), allowing \system to track which suggestions have been read.
The text spans used for generating suggestions are highlighted in light blue (\Cref{fig:interface}A); and when a user hovers over a suggestion in the Sidebar (\Cref{fig:interface}C), the corresponding text snippet is highlighted in yellow (Step 1 in \Cref{fig:interface}).

\paragraph{Four Ways of Generating Suggestions}
Writers using \system can select a portion of their draft and request suggestions from one of the following four different sources: 
\textbf{
{\em (i)} crowd workers (\textbf{\crowd}), 
{\em (ii)} a story plot generation model (\textbf{\plot}), 
{\em (iii)} GPT-3 that generates follow-up story arcs (\textbf{\gptplot}), and 
{\em (iv)} GPT-3 that completes the given story (\textbf{\gptc})}.
These are called suggestion ``task types'', which we detail as follows:
%
%
%
\textbf{{\em (i)} \crowd:}
The \crowd task allows the user to post tasks on Amazon Mechanical Turk to ask crowd workers to provide plot ideas that continue the selected story snippet.
Users can specify the number of ideas (the default is three) and write additional instructions for the workers.
We adapted the code from \oldSystem~\cite{heteroglossia-huang-2020} for the implementation.
\textbf{{\em (ii)} \plot:}
The \plot option generates a few sentences describing possible next events in the story based on the selected snippet.
This method 
predicts likely future events in a story based on its past~\cite{plotp2023huang}.
We adapted \citeauthor{plotp2023huang}'s data and code for the implementation.
\system offered two models for users to select:
one generates a \textit{near future} plot, summarizing what would happen in the next \textit{20} sentences;
the other generates a \textit{far future} plot, summarizing what would happen in the next \textit{100} sentences.
%
%
%
\textbf{{\em (iii)} \gptplot:}
We asked GPT-3~\cite{NEURIPS2020_1457c0d6} to produce a follow-up story arc (see Appendix~\ref{app:gpt-plot-promopt} for the prompt). 
%
%
%
\textbf{{\em (iv)} \gptc:}
This option continues the story based on the selected content by running GPT-3's text completion function (\texttt{text-davanci-003}).


\section{User Study}



\begin{table}[t]
\centering
\small
\begin{tabular}{@{}lllccccccccccc@{}}
\toprule
\multirow{2}{*}{\textbf{Part.}} & \multirow{2}{*}{\textbf{Writing Exp.}} & \multirow{2}{*}{\textbf{Publishing Exp.}} & \multirow{2}{*}{\textbf{Days}} & \multirow{2}{*}{\textbf{\#Word}} & \multicolumn{4}{c}{\textbf{\#Task}} & \multicolumn{4}{c}{\textbf{\#Suggestion}} \\ 
\cmidrule(lr){6-9}\cmidrule(lr){10-13}
& & & & & \textbf{C} & \textbf{S} & \textbf{GP} & \textbf{GC} & \textbf{C} & \textbf{S} & \textbf{GP} & \textbf{GC} \\ \hline
\participant{1} & $\leq$ 1 Year & Online forums, blogs, websites & 3 & 1,376 & \underline{1} & \textbf{3} & \textbf{3} & \textbf{3} & 1 & \underline{3} & \textbf{6} & \underline{3} \\
\participant{2} & 5-10 Years & Online forums, blogs, websites & 3 & 1,352 & 3 & \underline{4} & 2 & \textbf{5} & 5 & \underline{6} & 2 & \textbf{8} \\
\participant{4} & 5-10 Years & Published books & 7 & 2,376 & \underline{6} & \underline{6} & \underline{6} & \textbf{12} & \underline{6} & \underline{6} & \underline{6} & \textbf{12} \\
\participant{5} & $\geq$ 10 Years & Published books & 7 & 10,289 & \underline{6} & \textbf{11} & \underline{6} & \underline{6} & \textbf{12} & \underline{11} & 7 & 7 \\
\participant{10} & 5-10 Years & Online forums, blogs, websites & 7 & 1,773 & 1 & \underline{2} & 1 & \textbf{5} & \underline{2} & \underline{2} & \underline{2} & \textbf{9} \\
\participant{11} & 3-5 Years & Online forums, blogs, websites & 7 & 2,317 & 2 & 4 & \textbf{7} & \underline{5} & 2 & 6 & \textbf{9} & \underline{8} \\
\participant{14} & $\geq$ 10 Years & Online forums, blogs, websites & 7 & 2,715 & 5 & \underline{6} & 5 & \textbf{15} & \underline{11} & 10 & 8 & \textbf{16} \\
\participant{15} & $\geq$ 10 Years & Online forums, blogs, websites & 7 & 2,759 & 6 & \underline{9} & 7 & \textbf{10} & \underline{16} & \textbf{19} & 14 & \textbf{19} \\
\participant{2}$_{na}$ & 5-10 Years & Online forums, blogs, websites & 30 & - & 7 & \textbf{23} & 2 & \underline{13} & \underline{22} & \textbf{23} & 2 & 13 \\ \hline
\textbf{Total} & - & - & - & - & 37 & \underline{68} & 39 & \textbf{74} & 77 & \underline{86} & 56 & \textbf{95} \\
\bottomrule
\end{tabular}
\caption{Participant information and statistics. Part.: Participant, Writing Exp.: Writing Experience, Publishing Exp.: Publishing Experience, C: \crowd, S: \plot, GP: \gptplot, GC: \gptc. \participant{1} and \participant{2} joined the pilot study; the rest joined the formal user study. \participant{2}$_{na}$ represents \participant{2}'s NaNoWriMo study. The \textbf{highest} and \underline{second highest} usage are highlighted.}
\label{tab:participant-and-statistic}
\vspace{-9mm}
\end{table}

We developed a proof-of-concept system, \system, and conducted a deployment study to examine how writers interact with the system when given the option to use either crowd workers or AI. 
The study combined quantitative and qualitative approaches: user behavior was analyzed through detailed log data, and these findings were complemented by qualitative insights gathered through interviews with participants. 

\paragraph{Participants.}
\Cref{tab:participant-and-statistic} overviews all the participants.
For the pilot study, we recruited two participants through the authors' networks. 
In the main study, we expanded our recruitment efforts through social media posts, mailing lists at the authors' institute, and personal outreach.
Out of the 37 people who expressed interest, participants were selected based on two primary criteria. 
First, we gave preference to those with experience in publishing books, either through traditional publishing companies or self-publishing platforms. 
We were aware that most people in our recruiting pool were on the unpublished end of the experience spectrum, so prioritizing published authors helped ensure we captured a range of writing expertise. 
Second, we selected participants based on their writing process, distinguishing between those who outlined their stories before writing (outliners/architects), those who developed their stories as they wrote (discovery writers/brainstorms/gardeners), and those with a hybrid approach~\cite{reid1984radical}. 
This diversity in writing processes ensured a broader representation of writing practices in our participant pool.
The study was IRB-approved.

\paragraph{Writing Task.}
In the study, we asked each participant to use \system to write short stories within the specified duration (\eg, 7 days.)
The final outcome could be part of a longer story or a combination of short stories and did not need to be complete. 
Although we encouraged participants to meet a suggested word count (\eg, 2,000 words), it was not mandatory.
We also encouraged them to use each of the four suggestion-generation functions daily. 
We recorded all user activities in \system, including keystrokes, mouse clicks, and timestamps for creating and reading inspiration tasks.

\paragraph{Two-Part User Study.}
Our study had two parts, both of which used \system: a pilot study and the main study.
%
%
\textbf{(1) Pilot Study (3 Days, 2 Participants):}
The pilot study was 3-day-long, where the participants were encouraged to write 1,000 words.
The compensation for each participant was a \$100-dollar Amazon gift card.
All other setups of the pilot study were identical to the main study.
The pilot study led to final adjustments in \system.
\textbf{(2) Main User Study (7 Days, 6 Participants):}
The main study lasted 7 days, and participants were encouraged to write 2,000 words. 
We provided a \$200 Amazon Gift Card as incentive.
%
In both studies, on their first day using \system, participants completed a self-serve onboarding tutorial.
After each study, we conducted a 30 to 60-minute one-on-one semi-structured interview with each participant to gather feedback on \system. 
These interviews were audio-recorded for further analysis. Outside of the main user study, we conducted an extended case study with one participant (\participant{2}), who opted to continue use of \system during National Novel Writing Month (NaNoWriMo), which gave us an opportunity to study \system longitudinally (\ie, one month).
The goal of NaNoWriMo is to write a 50,000-word novel in November, aiming for a complete draft rather than perfection.
After the main study, the participant used \system for 30 more days.
We sent weekly check-in emails to ensure consistent usage.
Since participants preferred their familiar editor for NaNoWriMo and only used \system for suggestions,
we focused on qualitative insights from the semi-structured interview instead of analyzing writing logs.
We sent an additional \$200 Amazon Gift Card as compensation.

\section{Analyses and Findings}
We employed a mixed-methods approach, combining quantitative and qualitative analyses to address our three research questions (RQs).
To gather supportive evidence, we examined user behavioral and writing logs.
Additionally, we transcribed all interviews and conducted analytic induction,
a method that blends deductive and inductive approaches~\cite{robinson1951logical,znaniecki1934method}.
Our initial analysis of the interview data yielded 287 unique quotes or phrases from participants.
We developed a codebook and applied it to categorize 67 distinct excerpts.
These excerpts were then systematically grouped to address our three RQs.
Note that as the pilot study proved successful, we merged its data with the main study for analysis.
The NaNoWriMo case was excluded from the writing log analysis as the participant did not fully write on \system.








\begin{figure}[t]
    \centering
    \begin{minipage}{0.48\textwidth}
        \centering
        \small
        \addtolength{\tabcolsep}{-0.5mm}
        \begin{tabular}{@{}cccccc@{}}
        \toprule
        \multicolumn{2}{c}{\textbf{Task Type}} & \textbf{C} & \textbf{S} & \textbf{GP} & \textbf{GC} \\ \hline
        \multirow{4}{*}{\textbf{\begin{tabular}[c]{@{}c@{}}System\\ Latency\end{tabular}}} & \textbf{\#Samples} & 30 & 45 & 37 & 61 \\
         & \textbf{25\%} & 1,753 & 7 & 7 & 9 \\
         & \textbf{50\%} & 3,449 & 8 & 8 & 9 \\
         & \textbf{75\%} & 8,160 & 10 & 9 & 10 \\ \hline
        \multirow{4}{*}{\textbf{\begin{tabular}[c]{@{}c@{}}Reading\\ Latency\end{tabular}}} & \textbf{\#Samples} & 19 & 36 & 32 & 51 \\
         & \textbf{25\%} & 21,881 & 15 & 8 & 4 \\
         & \textbf{50\%} & 34,429 & 115 & 20 & 13 \\
         & \textbf{75\%} & 73,652 & 538 & 74 & 84 \\ \hline
        \multicolumn{2}{c}{\textbf{Not Reading}} & 36.67\% & 20.00\% & 13.51\% & 16.39\% \\ \bottomrule
        \end{tabular}
        \addtolength{\tabcolsep}{0.5mm}
        \captionof{table}{System and reading latency (in seconds) for four different tasks (RQ1).
            C: \crowd, S: \plot, GP: \gptplot, and GC: \gptc.
            Note that \participant{2}$_{na}$ is excluded. For \crowd, half took up to 60 minutes for an initial response, and users typically read the response hours later. A higher rate of not reading is also observed.}
        \label{tab:latency-distribution}
    \end{minipage}%
    \hfill
    \begin{minipage}{0.48\textwidth}
        \centering
        \vspace{-6mm}
        \includegraphics[width=0.95\linewidth]{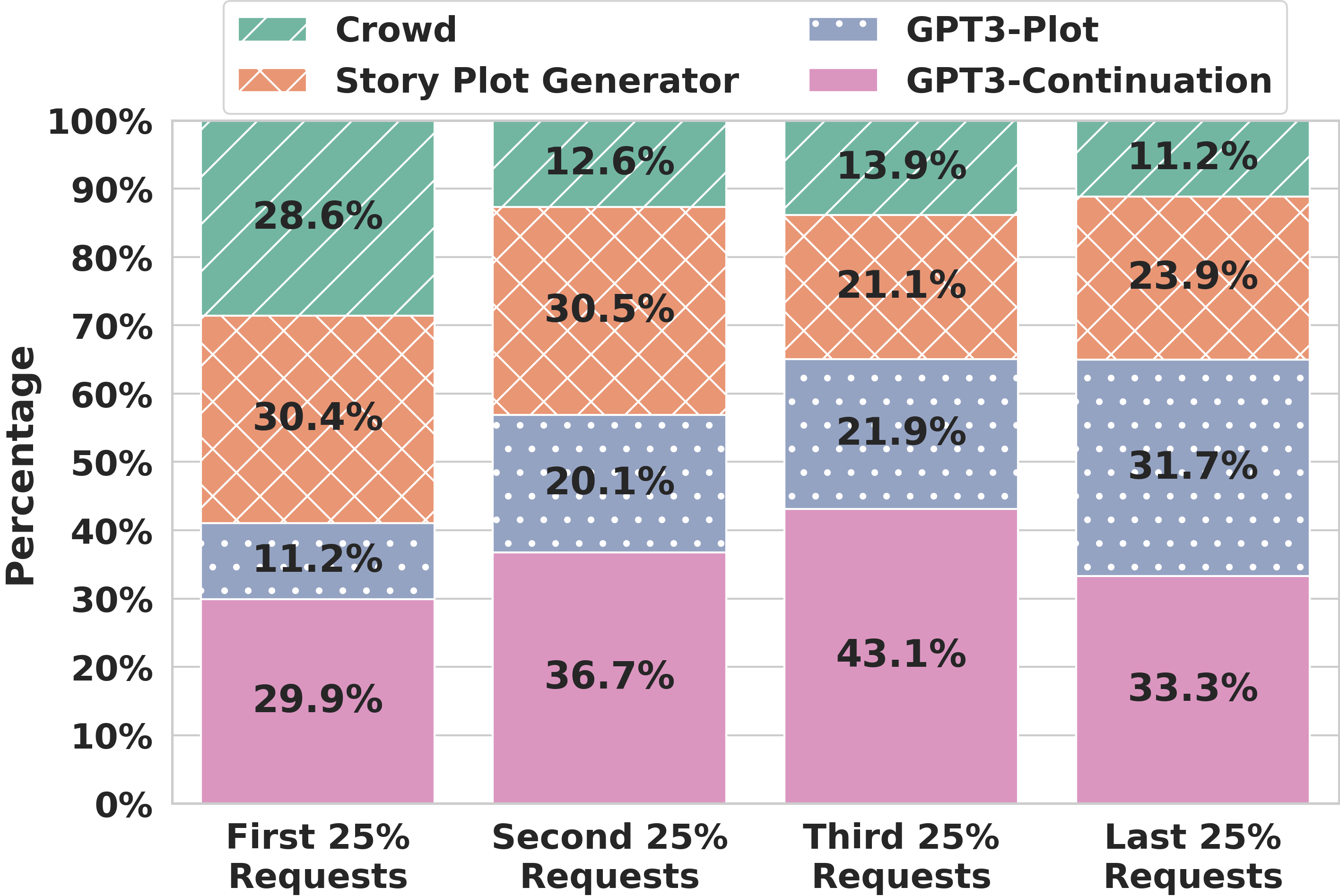}
        \vspace{-3mm}
        \captionof{figure}{Usage distribution of four functions across study phases, highlighting a decline in the \crowd function from 28.6\% initially to 11\%-13\% in subsequent phases (RQ1).}
        \label{fig:request-trend}
    \end{minipage}
    \vspace{-8mm}
\end{figure}

\subsection{RQ1: How frequently is each suggestion generation method used?}
\Cref{tab:participant-and-statistic} shows the usage statistics of the whole study.
\participant{5} incorporated their original works into the writing task, resulting in an extensive final draft of 10,287 words.
In contrast, \participant{15} produced three shorter drafts, with word counts of 1,198, 1,288, and 272, respectively.
Since the NaNoWriMo case did not require the participant to write within the \system's editor, we excluded the word counts for \participant{2}$_{na}$.

\subsubsection{\gptc was the most popular, while \crowd usage declined after the initial phase.}
\Cref{tab:participant-and-statistic} shows that the \gptc function was the most popular, with 74 tasks created and 95 suggestions generated. 
Six participants used \gptc most frequently.
Meanwhile, \plot was the second most popular option, primarily used by two participants (\participant{4} and \participant{2}$_{na}$).
We further analyzed the percentage of each function's use within the first, second, third, and last 25\% of participants' requests to \system.
Figure~\ref{fig:request-trend} shows that participants reduced their use of the \crowd option after the initial phase of the study. 
Initially, the crowd function accounted for 28.6\% of requests, but this dropped to 11\%-13\% in subsequent quarters.

From interviews with participants, we found that \textbf{all participants unanimously preferred \gptc}.
They appreciated the immediate responses provided by \gptc, which offered ``instant gratification'' (\participant{2})
and helped them overcome writer's block by generating fresh ideas.
Some participants mentioned that \gptc sometimes proposed alternative leads for completion, sparking new directions in their writing.
\participant{10} noted, \emph{``(\gptc) actually worked and worked very perfectly.''}
Some participants reported that \gptc disrupted their writing flow by prompting them to edit the AI-generated text, interrupting their thought process. 
Despite this, many still preferred \gptc over \crowd for its faster, immediate results.


On the other hand, participants valued the human touch and originality in the \crowd inputs,
expressing appreciation for the personal touch and shared understanding humans brought to their writing.
For example, \participant{14} mentioned that \emph{``everything was barely off topic... it made logical sense, any of the ideas or the continued story that was given. I think it was the most helpful overall.''}
Participants also enjoyed the ability to ask more directed questions and explore different alternatives for their writing.
Nevertheless, participants faced significant issues with inconsistent and delayed responses from the \crowd function.
The latency in receiving suggestions from crowd workers was a major deterrent.
As \participant{4} noted, \emph{``the timeframe was just... it's hard to not compare it to the other ones, the other ones were just so much faster. And so that was like the big drawback.''}
Participants suggested recruiting responders from diverse time zones to ensure continuous availability and improve timeliness.
Additionally, some participants found that the \crowd function sometimes produced incoherent responses due to unclear instructions.
They recommended providing clearer guidance for both users and crowd workers, as well as implementing mechanisms to verify the qualifications of human assistants.

\subsubsection{\crowd had a much longer system and reading latency, with one-third of responses going unread.}
From interviews with participants, we learned that latency significantly influenced their choices.
To analyze this, we calculated two types of latency for each function in \system: {\em (i)} system latency and {\em (ii)} reading latency.
System latency is the time from when a user creates a request to when the system first responds.
Reading latency measures the time from receiving the first response to when the user reads it.
Table~\ref{tab:latency-distribution} shows the results.
Most AI functions delivered responses within 10 seconds; users typically read these suggestions immediately.
In contrast, the \crowd function could take up to an hour for an initial response, and users often delayed reading these for several hours.
Moreover, participants did not read 36.67\% of the suggestions produced by the \crowd function.
Participants noted that the delays and inconsistent response times made the \crowd function less appealing compared to the immediate responses from AI features. 

\subsection{RQ2: How does each method of suggestion affect word count?}
For a comprehensive analysis of writing logs, we pre-processed the data as follows:
\textbf{{\em (i)} Step 1: Reconstructing Texts at Any Timestamp.}
We reconstructed the participants' writing at time \texttt{T} by collecting all \texttt{text-change} events occurring before time \texttt{T}. 
Each event includes a keystroke and a cursor offset. 
This set of \texttt{text-change} events was then rendered into a Delta document using \texttt{quill-delta}~\cite{quill-delta}.
Next, we converted the Delta document to HTML using \texttt{Quill-delta-python}~\cite{quill-delta-python} and extracted the text with \texttt{html2text}~\cite{html2text}.
\textbf{{\em (ii)} Step 2: Recreating Drafts Before and After Reading suggestions.}
We aim to measure the impact of suggestion on writing behavior, specifically after the writer reads the suggestion.
We extracted ``suggestion-reading events'' where participants interacted with the suggestion panel's tab by clicking and opening it. 
We then reconstructed and compared the state of the draft at two specific times: the moment of accessing the suggestion panel (time of the suggestion-reading event) and 300 seconds (5 minutes) thereafter.
We also prepared a comparison for 180 seconds (3 minutes).
\textbf{{\em (iii)} Step 3: Defining Working Sessions.}
We aim to simulate ``baseline'' writing behavior, helping us distinguish between short pauses for thinking and actual breaks from the task.
We followed a prior work~\cite{10.1145/3173574.3174023} to identify these sessions.
We first compiled all activity logs, including text changes and suggestion-reading events, and divided them into working sessions using an ``inactivity threshold.'' 
After testing thresholds ranging from 1 to 20 minutes, we recorded the number of sessions for each threshold.
Using the elbow rule~\cite{james2013introduction}, we identified 240 seconds as the optimal threshold,
which gave us 171 working sessions with a mean duration of 609 seconds (SD=973) and a median duration of 257 seconds.

\textbf{Baseline:}
RQ2 examines whether \system suggestions impact word count, 
a rough yet commonly used measure of writing progress. 
We used NLTK~\cite{bird2009natural} to calculate the document's word count when the suggestion panel was accessed and again 5 (or 3) minutes later.
To establish a baseline for writing behavior without \system suggestions,
we ran the following simulation 1,000 times: 
We randomly selected a working session, with the probability of selection proportional to the session's length. 
We then randomly picked a starting point within this session and measured the document changes 5 (or 3) minutes later.
Figure~\ref{fig:progress-distribution} shows KDE plots comparing the four suggestion functions with the baseline.

\subsubsection{\gptc boosted word production the most; \crowd might make writers think longer.}
\begin{figure}[t]
    \centering
    \begin{minipage}{0.496\textwidth}
        
        \centering
        \begin{subfigure}{0.95\linewidth}
            \centering
            \includegraphics[width=1.0\textwidth]{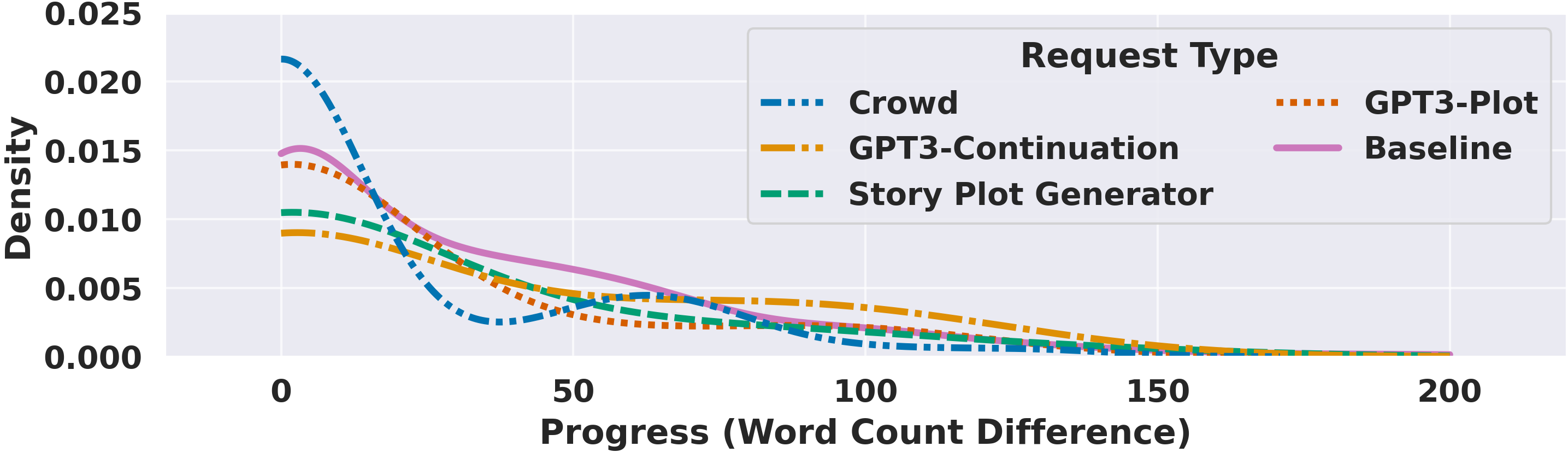}
            \vspace{-6mm}
            \caption{5-minute writing}
            \label{fig:progress-distribution-5-min}
        \end{subfigure}
        \begin{subfigure}{0.95\linewidth}
            \centering
            \includegraphics[width=1.0\textwidth]{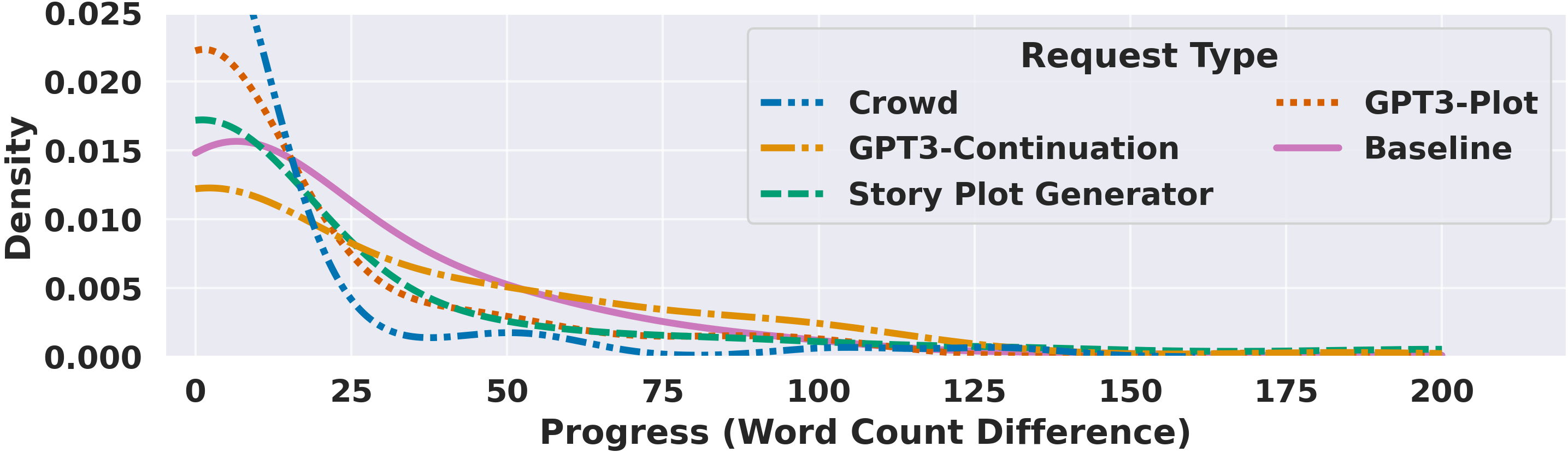}
            \vspace{-6mm}
            \caption{3-minute writing}
            \label{fig:progress-distribution-3-min}
        \end{subfigure}
        \vspace{-4mm}
        \caption{Distribution of the writing progress (RQ2). In (a), \gptc shows a higher density near 100 words; \crowd comes with a small peak at 70 words. This peak reduces and shifts left in (b), indicating that users might need time to catch up with the prior information. (We set \texttt{common\_norm} to \texttt{false} to ensure each distribution's area sums to one. We also clipped values to $[0, 200]$.)}
        \label{fig:progress-distribution}
        
    \end{minipage}
    \hfill
    \begin{minipage}{0.485\textwidth}
        
        \centering
        \includegraphics[width=1.0\linewidth]{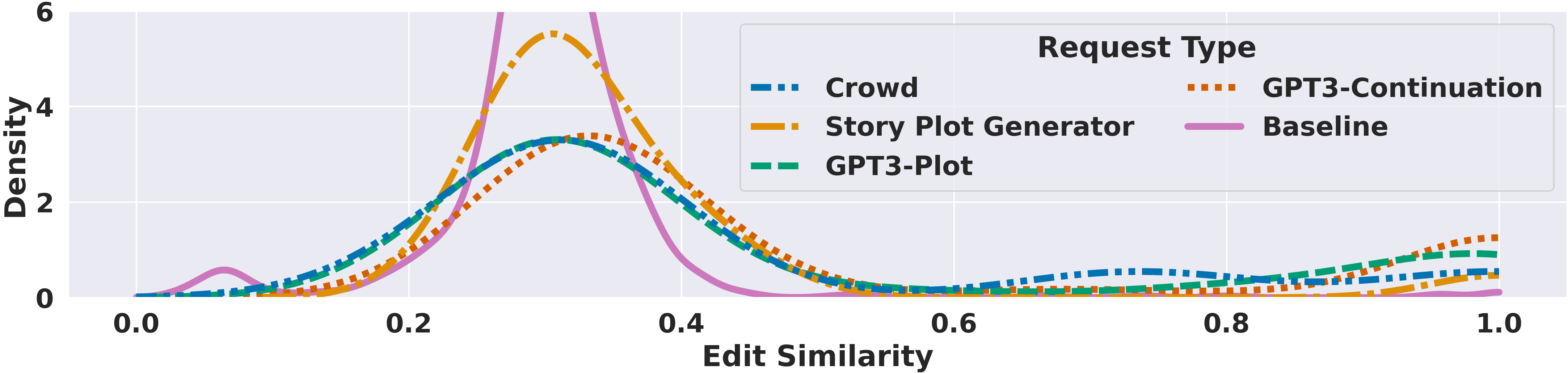}
        \hfill
        \includegraphics[width=1.0\linewidth]{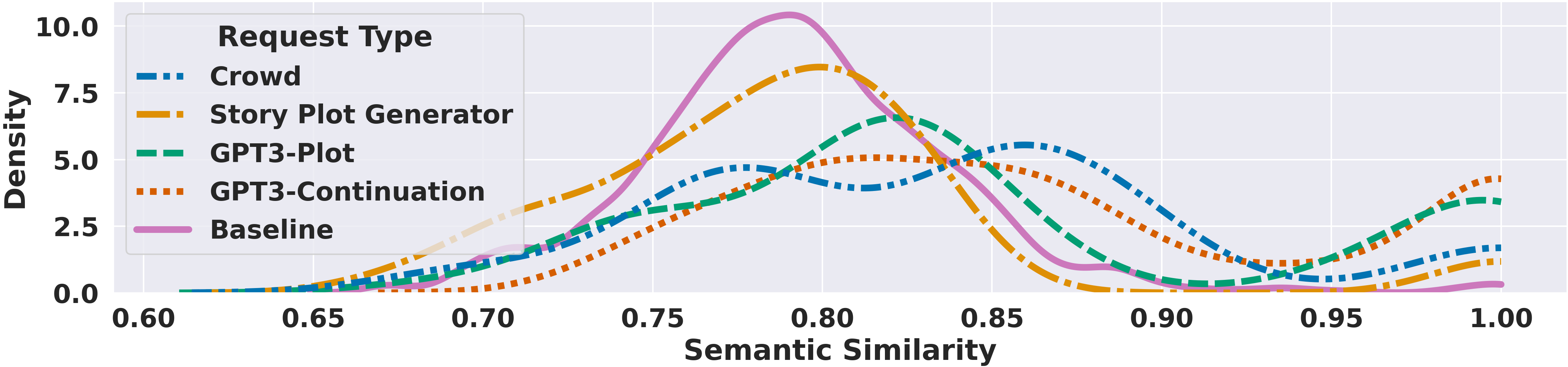}
        \includegraphics[width=1.0\linewidth]{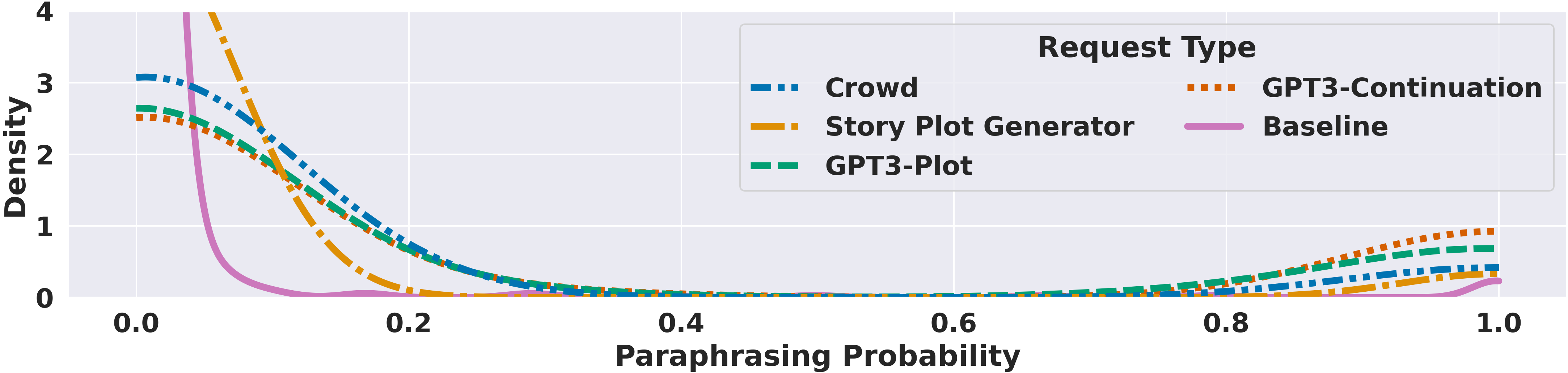}
        \hfill
        \vspace{-4mm}
        \caption{Distribution of the similarity metrics (RQ3). A 1.0 similarity means that the whole sentence is adopted. Overall, \gptc has a higher rate of being adopted.}
        \label{fig:similarity-behavior}
        \vspace{-2mm}
        
    \end{minipage}
    \vspace{-5mm}
\end{figure}

In \Cref{fig:progress-distribution-5-min}, the distribution from \gptc is flatter than all other options and the baseline.
\gptc's main peak around 0 words---indicating no additional words produced---is also the lowest among all options. 
These suggest that \gptc is the most effective in boosting word production.
\plot and \gptplot, while not as effective as \gptc, still made participants produce more words than the baseline.
We performed the same analysis with a 3-minute window (\Cref{fig:progress-distribution-3-min}), \gptc remains superior.

The \crowd function shows the highest main peak at 0 words, suggesting participants often did not produce any words within 5 minutes of \textit{reading} its suggestion. 
We hypothesize that this occurs because \crowd suggestions frequently relate to drafts written hours earlier, requiring users to spend time reacquainting themselves with the context before continuing their writing.
Furthermore, a more pronounced peak at 0 words in a 3-minute analysis window (\Cref{fig:progress-distribution-3-min}) indicates that participants were even less likely to incorporate \crowd suggestion within 3 minutes, and slightly more likely to do so within 5 minutes.
Further quantitative research and data are needed to fully understand these phenomena.

Our interviews show more nuanced details on how different suggestion-generation methods influence participants' writing process. 
The immediate availability of \gptc suggestions appears to contribute to its popularity and effectiveness in boosting word production. 
For example, participants frequently used \gptc for ``text dumps,'' allowing them to insert generic ideas for later revision without needing to articulate precise thoughts on the story. 
As one participant explained, \emph{``There were sometimes pieces of information that I could pull from it.'' } (\participant{15}).
In contrast, while the delayed nature of \crowd suggestions could be frustrating at times, some participants appreciated the depth it added to their writing process. 
One participant said, \emph{``Sometimes I needed to see how what I'd written had come across before moving on to the next part, and I was able to get a response from an actual person. Like an actual reader, essentially, to see if what I was saying made sense to them or what they thought was happening.''} (\participant{14}).

\subsection{RQ3: How does each suggestion method influence writing content?}
We compared sentences in the suggestions to the ``newly-edited content'' in the drafts. 
``Newly-edited content'' includes {\em (i)} sentences that were not in the draft when the suggestion was accessed but appeared five minutes later, and  {\em (ii)} sentences that were edited within these five minutes.
For each suggestion-reading event (see RQ2), we first calculated sentence-level similarities pairwisely.
For example, with a 3-sentence suggestion and a draft having 5-sentence newly-edited content, we calculated 15 ($3\times5$) similarities. 
We then took the highest value among the 15 as the final similarity score,
considering that even one closely matching sentence indicates a strong influence~\cite{roemmele2021inspiration}.
We applied three sentence-level similarity metrics:
\textbf{{\em (i)} Normalized Edit Similarity:} $1 - Normalized Edit Distance$~\cite{4160958,python-string-similarity}, where higher scores indicate higher similarity;
\textbf{{\em (ii)} Semantic Similarity:}
cosine similarity on semantic representations obtained from Sentence BERT~\cite{reimers-2019-sentence-bert};
and \textbf{{\em (iii)} Paraphrasing Detection Score:}
an output probability from a paraphrasing detection model~\cite{nighojkar-licato-2021-improving} to determine if a written sentence is a paraphrase of one of the suggestions.

\textbf{Baseline:}
We simulated a baseline condition by comparing a random working draft and a random suggestion. 
We randomly selected a time from a working session and observed changes five minutes later.
We also randomly sampled one suggestion from the same document and then applied the three similarity metrics to these sentences. 
The simulation was run 1,000 times.
Finally, similar to RD2, we used KDE plots to visualize the influence levels across the four suggestion methods and the baseline condition.
Figure~\ref{fig:similarity-behavior} shows the results, in which
a 1.0 similarity indicates complete integration of the ``suggestion sentence'' into the user's draft, while a 0.0 similarity means no integration occurred.
In each chart, the prominent Baseline peak represents cases where users did not adopt any parts of the generated suggestion.

\subsubsection{\gptc influenced the written content the most, with full-sentence adoption across all methods.}
Across all three charts in Figure~\ref{fig:similarity-behavior}, \gptc's curves are flatter and skewed to the right, indicating a higher likelihood of users producing texts similar to those generated by \gptc.
Additionally, at the high-similarity end (close to 1.0), \gptc exhibits a greater probability density compared to other options. 
This indicates a higher tendency for \gptc sentences to be directly copied into participants' drafts. 
Overall, \gptc influences the participants' writing the most.
Note that, at the high-similarity ends, all options exhibit a probability density greater than zero, indicating that full-sentence adoption occurs across all suggestion methods in \system.

From the interviews, participants confirmed that \gptc often directly influenced their writing content,
with some participants incorporating sentences or phrases directly from the AI-generated suggestions into their drafts.
They found \gptc helpful in providing immediate continuations that aligned closely with their writing style and narrative.
Participants had mixed feelings about using \gptplot and \plot functions.
Some used \gptplot to validate their storylines or explore alternative perspectives,
appreciating its quick response time and utility in mapping out structural outlines.
\participant{4} mentioned, \emph{``the function was helpful for figuring out how to map out a plot. I didn't really start with a plot in mind when I was writing this story.''}
However, others found that \gptplot and \plot sometimes produced irrelevant suggestions due to a poor grasp of context or failed to provide refined guidance in specific scenarios.
\participant{14} noted, \emph{``I stopped using the near [future option] after a couple of days because I wasn't getting anything helpful from it. So after I learned that the distant [future option] was usually more helpful for me.''}

Participants also noted that the AI sometimes struggled with complex prompts and did not always align with their expectations,
which could limit the influence of these suggestions on their writing content.
They expressed a desire for more interactive AI experiences,
including the ability to ask follow-up questions and receive meaningful responses.

\subsubsection{Users sometimes seek validation from \system.}
\label{sec:soc-findings}
Interviews with participants and observations showed that, even though participants were aware of their own intent, they sometimes turned to \system for validation of their thoughts.
%
%
\participant{5} explained this process, \emph{``I asked (\system), should she confess the lie to her brother? Or should she insist that there is a boyfriend? And \textbf{it (\system) actually got right on what I was going to do} anyhow... Sometimes this could be either \textbf{confirmation of what I was going to do} or kind of lead me in another direction.''}
\participant{15} also validate the story's trajectory, \emph{``I want to see what things, like where they (\system) think this story is going and then see \textbf{whether or not that validates where I have it going in mind.''}}
%
%
This validation-seeking behavior occurs much more frequently with AI options than with \crowd options, perhaps due to the faster response times of AI. 
A system with a similar design to ours but relying solely on crowd workers without AI, \oldSystem~\cite{heteroglossia-huang-2020}, did not report this behavior.

In some cases, participants expressed frustration when the system failed to align with the norms or assumptions in their writing. 
For example, \participant{1}, who was writing a queer story, noted that the AI consistently suggested heterosexual relationship dynamics and stereotypical portrayals:
\emph{``I could imagine some people if they're writing a \textbf{queer story}... \textbf{all the suggestions that (\system) gave me were heterosexual relationships}.''} and \emph{``... (\plot) gave me something like `she blushed when she realized people were looking at her naked body.'... like women are dispensable.''}

\section{Discussion and Design Suggestions}



Our study suggests that, with the increasing accessibility of LLMs, it is probably no longer necessary to build crowd-writing systems solely to explore how powerful text generation and language understanding components can enhance writing. 
Instead, the other goal of crowd-writing systems---examining how smaller, well-defined contributions from human workers can support writing---remains highly relevant. 
As humans continue to outperform AI in creative writing, understanding how to effectively integrate human input alongside AI is becoming even more crucial.
Based on our study, we propose three design suggestions (DS):

\textbf{DS \#1: Account for the differences between human and AI assistance.}
\label{}
Our first suggestion is to design the system in a way that clearly accounts for the differences between human assistance and AI assistance. 
In our study, the crowd assistance was significantly different from the AI options. 
The most obvious disparity was their response speed; participants also noted greater variability in the quality of the crowd assistance, while mentioning that the crowd options sometimes offered a more ``human touch.'' 
However, our study followed the design of \oldSystem, which delivered both AI and human assistance through the same modality and interface.
This approach proved ineffective. 
We, therefore, advocate for a system design that better highlights and accommodates these differences.

\textbf{DS \#2: Design against automation bias, especially when both human and AI assistance were presented.}
Our study shows that when AI
is presented as an option alongside crowdsourcing, users tend to favor AI (RQ1). 
As a result, AI has a more direct influence on the writing process: more words are generated based on AI suggestions (RQ2), and more of the content is shaped semantically by AI's output (RQ3). 
Given that humans still outperform AI in creative writing---as shown in a series of literature~\cite{bellemare2024divergent}---this influence may not always be desired.
Therefore, our second design implication is to design against automation bias---the human tendency to over-rely on or accept machine-generated outputs~\cite{goddard2012automation,10.1136/amiajnl-2011-000089}. 
Designers of writing systems must be mindful of this bias and work to mitigate the potential negative impact of AI on users' writing.
This becomes particularly important if the goal is to enable users to benefit from both human and automated assistance, as it is often much easier for users to rely on AI-generated quick suggestions.

\textbf{DS \#3: Consider the \soc view of writing (in addition to the commonly used cognitive process view).}
We advocate for a \textbf{\soc view of writing} into the design of writing assistants, alongside the already widely-used cognitive views~\cite{flower1981cognitive,greer2016introduction,feldman2021we}.
\Soc theories of writing offer a holistic interpretation of users' validation-seeking behaviors, as well as the frustration that may arise when these behaviors clash with the AI's expected norms---something that cognitive process theory cannot fully explain.
Discourse Community (DC) is a \soc concept that views a genre of writing as a constantly evolving and shared form of communication for a group of individuals who share values, interests, or goals~\cite{killingsworth1992,swales2017concept,Wenger2000}. 
Writers shape their work based on their discourse community's expectations; 
community members understand unspoken rules and expectations~\cite{becker2000beyond}, often motivating writers in ways that do not need to be explicitly stated. 
Our findings suggest that writers sometimes used \system's generated text to double-check their alignment with these expectations, which raises several potential risks. 
First, LLM outputs, at best, approximate community conventions rather than reflecting the actual, evolving norms~\cite{Wenger2000}. 
More critically, LLMs may reinforce outdated voices within a discourse community while suppressing emerging ones, as seen with \participant{1} (Section~\ref{sec:soc-findings}). 
Like many technologies, LLMs risk disproportionately benefiting those already in positions of power.

\section{Conclusion and Future Work}
This paper introduces \system, a text editor that allows creative writers to request inspiration from online crowd workers and three different AI models, including  GPT-3 for continuing ongoing stories (\gptc).
A week-long user study with multiple creative writers showed that participants consistently decreased their use of crowds in favor of AIs throughout the study. 
Motivated by the study's observations, we offer a set of design suggestions for future crowd-AI hybrid writing systems.
Future work will focus on conducting larger-scale studies with expanded \system features and making improvements to model outputs. 

\bibliographystyle{ACM-Reference-Format}
\bibliography{bib/aaai22,bib/c-and-c24}


\begin{thebibliography}{41}


\ifx \showCODEN    \undefined \def \showCODEN     #1{\unskip}     \fi
\ifx \showDOI      \undefined \def \showDOI       #1{#1}\fi
\ifx \showISBNx    \undefined \def \showISBNx     #1{\unskip}     \fi
\ifx \showISBNxiii \undefined \def \showISBNxiii  #1{\unskip}     \fi
\ifx \showISSN     \undefined \def \showISSN      #1{\unskip}     \fi
\ifx \showLCCN     \undefined \def \showLCCN      #1{\unskip}     \fi
\ifx \shownote     \undefined \def \shownote      #1{#1}          \fi
\ifx \showarticletitle \undefined \def \showarticletitle #1{#1}   \fi
\ifx \showURL      \undefined \def \showURL       {\relax}        \fi
\providecommand\bibfield[2]{#2}
\providecommand\bibinfo[2]{#2}
\providecommand\natexlab[1]{#1}
\providecommand\showeprint[2][]{arXiv:#2}

\bibitem[Ackerman(2000)]%
        {ackerman2000intellectual}
\bibfield{author}{\bibinfo{person}{Mark~S Ackerman}.} \bibinfo{year}{2000}\natexlab{}.
\newblock \showarticletitle{The intellectual challenge of CSCW: the gap between social requirements and technical feasibility}.
\newblock \bibinfo{journal}{\emph{Human--Computer Interaction}} \bibinfo{volume}{15}, \bibinfo{number}{2-3} (\bibinfo{year}{2000}), \bibinfo{pages}{179--203}.
\newblock


\bibitem[Alir3z4(2024)]%
        {html2text}
\bibfield{author}{\bibinfo{person}{Alir3z4}.} \bibinfo{year}{2024}\natexlab{}.
\newblock \bibinfo{title}{html2text}.
\newblock
\newblock
\urldef\tempurl%
\url{{https://github.com/Alir3z4/html2text}}
\showURL{%
\tempurl}


\bibitem[Becker(2000)]%
        {becker2000beyond}
\bibfield{author}{\bibinfo{person}{Alton~L Becker}.} \bibinfo{year}{2000}\natexlab{}.
\newblock \bibinfo{booktitle}{\emph{Beyond translation: Essays toward a modern philology}}.
\newblock \bibinfo{publisher}{University of Michigan Press}.
\newblock


\bibitem[Begus(2023)]%
        {begus2023experimental}
\bibfield{author}{\bibinfo{person}{Nina Begus}.} \bibinfo{year}{2023}\natexlab{}.
\newblock \showarticletitle{Experimental Narratives: A Comparison of Human Crowdsourced Storytelling and AI Storytelling}.
\newblock \bibinfo{journal}{\emph{arXiv preprint arXiv:2310.12902}} (\bibinfo{year}{2023}).
\newblock


\bibitem[Bellemare-Pepin et~al\mbox{.}(2024)]%
        {bellemare2024divergent}
\bibfield{author}{\bibinfo{person}{Antoine Bellemare-Pepin}, \bibinfo{person}{Fran{\c{c}}ois Lespinasse}, \bibinfo{person}{Philipp Th{\"o}lke}, \bibinfo{person}{Yann Harel}, \bibinfo{person}{Kory Mathewson}, \bibinfo{person}{Jay~A Olson}, \bibinfo{person}{Yoshua Bengio}, {and} \bibinfo{person}{Karim Jerbi}.} \bibinfo{year}{2024}\natexlab{}.
\newblock \showarticletitle{Divergent Creativity in Humans and Large Language Models}.
\newblock \bibinfo{journal}{\emph{arXiv preprint arXiv:2405.13012}} (\bibinfo{year}{2024}).
\newblock


\bibitem[Bernstein et~al\mbox{.}(2010)]%
        {bernstein2010soylent}
\bibfield{author}{\bibinfo{person}{Michael~S Bernstein}, \bibinfo{person}{Greg Little}, \bibinfo{person}{Robert~C Miller}, \bibinfo{person}{Bj{\"o}rn Hartmann}, \bibinfo{person}{Mark~S Ackerman}, \bibinfo{person}{David~R Karger}, \bibinfo{person}{David Crowell}, {and} \bibinfo{person}{Katrina Panovich}.} \bibinfo{year}{2010}\natexlab{}.
\newblock \showarticletitle{Soylent: a word processor with a crowd inside}. In \bibinfo{booktitle}{\emph{Proceedings of the 23nd annual ACM symposium on User interface software and technology}}. \bibinfo{pages}{313--322}.
\newblock


\bibitem[Bird et~al\mbox{.}(2009)]%
        {bird2009natural}
\bibfield{author}{\bibinfo{person}{Steven Bird}, \bibinfo{person}{Ewan Klein}, {and} \bibinfo{person}{Edward Loper}.} \bibinfo{year}{2009}\natexlab{}.
\newblock \bibinfo{booktitle}{\emph{Natural language processing with Python: analyzing text with the natural language toolkit}}.
\newblock \bibinfo{publisher}{" O'Reilly Media, Inc."}.
\newblock


\bibitem[Brown et~al\mbox{.}(2020)]%
        {NEURIPS2020_1457c0d6}
\bibfield{author}{\bibinfo{person}{Tom Brown}, \bibinfo{person}{Benjamin Mann}, \bibinfo{person}{Nick Ryder}, \bibinfo{person}{Melanie Subbiah}, \bibinfo{person}{Jared~D Kaplan}, \bibinfo{person}{Prafulla Dhariwal}, \bibinfo{person}{Arvind Neelakantan}, \bibinfo{person}{Pranav Shyam}, \bibinfo{person}{Girish Sastry}, \bibinfo{person}{Amanda Askell}, \bibinfo{person}{Sandhini Agarwal}, \bibinfo{person}{Ariel Herbert-Voss}, \bibinfo{person}{Gretchen Krueger}, \bibinfo{person}{Tom Henighan}, \bibinfo{person}{Rewon Child}, \bibinfo{person}{Aditya Ramesh}, \bibinfo{person}{Daniel Ziegler}, \bibinfo{person}{Jeffrey Wu}, \bibinfo{person}{Clemens Winter}, \bibinfo{person}{Chris Hesse}, \bibinfo{person}{Mark Chen}, \bibinfo{person}{Eric Sigler}, \bibinfo{person}{Mateusz Litwin}, \bibinfo{person}{Scott Gray}, \bibinfo{person}{Benjamin Chess}, \bibinfo{person}{Jack Clark}, \bibinfo{person}{Christopher Berner}, \bibinfo{person}{Sam McCandlish}, \bibinfo{person}{Alec Radford}, \bibinfo{person}{Ilya Sutskever}, {and}
  \bibinfo{person}{Dario Amodei}.} \bibinfo{year}{2020}\natexlab{}.
\newblock \showarticletitle{Language Models are Few-Shot Learners}. In \bibinfo{booktitle}{\emph{Advances in Neural Information Processing Systems}}, \bibfield{editor}{\bibinfo{person}{H.~Larochelle}, \bibinfo{person}{M.~Ranzato}, \bibinfo{person}{R.~Hadsell}, \bibinfo{person}{M.F. Balcan}, {and} \bibinfo{person}{H.~Lin}} (Eds.), Vol.~\bibinfo{volume}{33}. \bibinfo{publisher}{Curran Associates, Inc.}, \bibinfo{pages}{1877--1901}.
\newblock
\urldef\tempurl%
\url{https://proceedings.neurips.cc/paper/2020/file/1457c0d6bfcb4967418bfb8ac142f64a-Paper.pdf}
\showURL{%
\tempurl}


\bibitem[Chakrabarty et~al\mbox{.}(2024a)]%
        {10.1145/3613904.3642731}
\bibfield{author}{\bibinfo{person}{Tuhin Chakrabarty}, \bibinfo{person}{Philippe Laban}, \bibinfo{person}{Divyansh Agarwal}, \bibinfo{person}{Smaranda Muresan}, {and} \bibinfo{person}{Chien-Sheng Wu}.} \bibinfo{year}{2024}\natexlab{a}.
\newblock \showarticletitle{Art or Artifice? Large Language Models and the False Promise of Creativity}. In \bibinfo{booktitle}{\emph{Proceedings of the 2024 CHI Conference on Human Factors in Computing Systems}} (Honolulu, HI, USA) \emph{(\bibinfo{series}{CHI '24})}. \bibinfo{publisher}{Association for Computing Machinery}, \bibinfo{address}{New York, NY, USA}, Article \bibinfo{articleno}{30}, \bibinfo{numpages}{34}~pages.
\newblock
\showISBNx{9798400703300}
\urldef\tempurl%
\url{https://doi.org/10.1145/3613904.3642731}
\showDOI{\tempurl}


\bibitem[Chakrabarty et~al\mbox{.}(2024b)]%
        {chakrabarty2024can}
\bibfield{author}{\bibinfo{person}{Tuhin Chakrabarty}, \bibinfo{person}{Philippe Laban}, {and} \bibinfo{person}{Chien-Sheng Wu}.} \bibinfo{year}{2024}\natexlab{b}.
\newblock \showarticletitle{Can AI writing be salvaged? Mitigating Idiosyncrasies and Improving Human-AI Alignment in the Writing Process through Edits}.
\newblock \bibinfo{journal}{\emph{arXiv preprint arXiv:2409.14509}} (\bibinfo{year}{2024}).
\newblock


\bibitem[Chakrabarty et~al\mbox{.}(2024c)]%
        {10.1145/3635636.3656201}
\bibfield{author}{\bibinfo{person}{Tuhin Chakrabarty}, \bibinfo{person}{Vishakh Padmakumar}, \bibinfo{person}{Faeze Brahman}, {and} \bibinfo{person}{Smaranda Muresan}.} \bibinfo{year}{2024}\natexlab{c}.
\newblock \showarticletitle{Creativity Support in the Age of Large Language Models: An Empirical Study Involving Professional Writers}. In \bibinfo{booktitle}{\emph{Proceedings of the 16th Conference on Creativity \& Cognition}} (Chicago, IL, USA) \emph{(\bibinfo{series}{C\&C '24})}. \bibinfo{publisher}{Association for Computing Machinery}, \bibinfo{address}{New York, NY, USA}, \bibinfo{pages}{132–155}.
\newblock
\showISBNx{9798400704857}
\urldef\tempurl%
\url{https://doi.org/10.1145/3635636.3656201}
\showDOI{\tempurl}


\bibitem[Feldman and McInnis(2021)]%
        {feldman2021we}
\bibfield{author}{\bibinfo{person}{Molly~Q Feldman} {and} \bibinfo{person}{Brian~James McInnis}.} \bibinfo{year}{2021}\natexlab{}.
\newblock \showarticletitle{How we write with crowds}.
\newblock \bibinfo{journal}{\emph{Proceedings of the ACM on Human-Computer Interaction}} \bibinfo{volume}{4}, \bibinfo{number}{CSCW3} (\bibinfo{year}{2021}), \bibinfo{pages}{1--31}.
\newblock


\bibitem[Flower and Hayes(1981)]%
        {flower1981cognitive}
\bibfield{author}{\bibinfo{person}{Linda Flower} {and} \bibinfo{person}{John~R Hayes}.} \bibinfo{year}{1981}\natexlab{}.
\newblock \showarticletitle{A cognitive process theory of writing}.
\newblock \bibinfo{journal}{\emph{College composition and communication}} \bibinfo{volume}{32}, \bibinfo{number}{4} (\bibinfo{year}{1981}), \bibinfo{pages}{365--387}.
\newblock


\bibitem[forgeworks(2024)]%
        {quill-delta-python}
\bibfield{author}{\bibinfo{person}{forgeworks}.} \bibinfo{year}{2024}\natexlab{}.
\newblock \bibinfo{title}{quill-delta-python}.
\newblock
\newblock
\urldef\tempurl%
\url{{https://github.com/forgeworks/quill-delta-python}}
\showURL{%
\tempurl}


\bibitem[Goddard et~al\mbox{.}(2011)]%
        {10.1136/amiajnl-2011-000089}
\bibfield{author}{\bibinfo{person}{Kate Goddard}, \bibinfo{person}{Abdul Roudsari}, {and} \bibinfo{person}{Jeremy~C Wyatt}.} \bibinfo{year}{2011}\natexlab{}.
\newblock \showarticletitle{{Automation bias: a systematic review of frequency, effect mediators, and mitigators}}.
\newblock \bibinfo{journal}{\emph{Journal of the American Medical Informatics Association}} \bibinfo{volume}{19}, \bibinfo{number}{1} (\bibinfo{date}{06} \bibinfo{year}{2011}), \bibinfo{pages}{121--127}.
\newblock
\showISSN{1067-5027}
\urldef\tempurl%
\url{https://doi.org/10.1136/amiajnl-2011-000089}
\showDOI{\tempurl}
\showeprint{https://academic.oup.com/jamia/article-pdf/19/1/121/5911703/19-1-121.pdf}


\bibitem[Goddard et~al\mbox{.}(2012)]%
        {goddard2012automation}
\bibfield{author}{\bibinfo{person}{Kate Goddard}, \bibinfo{person}{Abdul Roudsari}, {and} \bibinfo{person}{Jeremy~C Wyatt}.} \bibinfo{year}{2012}\natexlab{}.
\newblock \showarticletitle{Automation bias: a systematic review of frequency, effect mediators, and mitigators}.
\newblock \bibinfo{journal}{\emph{Journal of the American Medical Informatics Association}} \bibinfo{volume}{19}, \bibinfo{number}{1} (\bibinfo{year}{2012}), \bibinfo{pages}{121--127}.
\newblock


\bibitem[Greer et~al\mbox{.}(2016)]%
        {greer2016introduction}
\bibfield{author}{\bibinfo{person}{Nick Greer}, \bibinfo{person}{Jaime Teevan}, {and} \bibinfo{person}{Shamsi~T Iqbal}.} \bibinfo{year}{2016}\natexlab{}.
\newblock \bibinfo{booktitle}{\emph{An introduction to technological support for writing}}.
\newblock \bibinfo{type}{{T}echnical {R}eport}. \bibinfo{institution}{Technical Report. Microsoft Research Tech Report MSR-TR-2016-001}.
\newblock


\bibitem[Hahn et~al\mbox{.}(2016)]%
        {hahn2016knowledge}
\bibfield{author}{\bibinfo{person}{Nathan Hahn}, \bibinfo{person}{Joseph Chang}, \bibinfo{person}{Ji~Eun Kim}, {and} \bibinfo{person}{Aniket Kittur}.} \bibinfo{year}{2016}\natexlab{}.
\newblock \showarticletitle{The Knowledge Accelerator: Big picture thinking in small pieces}. In \bibinfo{booktitle}{\emph{Proceedings of the 2016 CHI Conference on Human Factors in Computing Systems}}. \bibinfo{pages}{2258--2270}.
\newblock


\bibitem[Hara et~al\mbox{.}(2018)]%
        {10.1145/3173574.3174023}
\bibfield{author}{\bibinfo{person}{Kotaro Hara}, \bibinfo{person}{Abigail Adams}, \bibinfo{person}{Kristy Milland}, \bibinfo{person}{Saiph Savage}, \bibinfo{person}{Chris Callison-Burch}, {and} \bibinfo{person}{Jeffrey~P. Bigham}.} \bibinfo{year}{2018}\natexlab{}.
\newblock \showarticletitle{A Data-Driven Analysis of Workers' Earnings on Amazon Mechanical Turk}. In \bibinfo{booktitle}{\emph{Proceedings of the 2018 CHI Conference on Human Factors in Computing Systems}} (Montreal QC, Canada) \emph{(\bibinfo{series}{CHI '18})}. \bibinfo{publisher}{Association for Computing Machinery}, \bibinfo{address}{New York, NY, USA}, \bibinfo{pages}{1–14}.
\newblock
\showISBNx{9781450356206}
\urldef\tempurl%
\url{https://doi.org/10.1145/3173574.3174023}
\showDOI{\tempurl}


\bibitem[Huang et~al\mbox{.}(2020)]%
        {heteroglossia-huang-2020}
\bibfield{author}{\bibinfo{person}{Chieh-Yang Huang}, \bibinfo{person}{Shih-Hong Huang}, {and} \bibinfo{person}{Ting-Hao~Kenneth Huang}.} \bibinfo{year}{2020}\natexlab{}.
\newblock \showarticletitle{Heteroglossia: In-Situ Story Ideation with the Crowd}. In \bibinfo{booktitle}{\emph{Proceedings of the 2020 CHI Conference on Human Factors in Computing Systems}} (Honolulu, HI, USA) \emph{(\bibinfo{series}{CHI '20})}. \bibinfo{publisher}{Association for Computing Machinery}, \bibinfo{address}{New York, NY, USA}, \bibinfo{pages}{1–12}.
\newblock
\showISBNx{9781450367080}
\urldef\tempurl%
\url{https://doi.org/10.1145/3313831.3376715}
\showDOI{\tempurl}


\bibitem[Huang et~al\mbox{.}(2023)]%
        {plotp2023huang}
\bibfield{author}{\bibinfo{person}{Chieh-Yang Huang}, \bibinfo{person}{Saniya Naphade}, \bibinfo{person}{Kavya~Laalasa Karanam}, {and} \bibinfo{person}{Ting-Hao~K. Huang}.} \bibinfo{year}{2023}\natexlab{}.
\newblock \showarticletitle{Conveying the Predicted Future to Users: A Case Study of Story Plot Prediction}. In \bibinfo{booktitle}{\emph{Proc. AAAI 2023 Workshop of Creative AI Across Modalities}}.
\newblock


\bibitem[James et~al\mbox{.}(2013)]%
        {james2013introduction}
\bibfield{author}{\bibinfo{person}{Gareth James}, \bibinfo{person}{Daniela Witten}, \bibinfo{person}{Trevor Hastie}, \bibinfo{person}{Robert Tibshirani}, {et~al\mbox{.}}} \bibinfo{year}{2013}\natexlab{}.
\newblock \bibinfo{booktitle}{\emph{An introduction to statistical learning}}. Vol.~\bibinfo{volume}{112}.
\newblock \bibinfo{publisher}{Springer}.
\newblock


\bibitem[Killingsworth(1992)]%
        {killingsworth1992}
\bibfield{author}{\bibinfo{person}{M.~Jimmie Killingsworth}.} \bibinfo{year}{1992}\natexlab{}.
\newblock \showarticletitle{Discourse Communities. Local and Global}.
\newblock \bibinfo{journal}{\emph{Rhetoric Review}} \bibinfo{volume}{11}, \bibinfo{number}{1} (\bibinfo{year}{1992}), \bibinfo{pages}{110--122}.
\newblock
\showISSN{07350198, 15327981}
\urldef\tempurl%
\url{http://www.jstor.org/stable/465883}
\showURL{%
\tempurl}


\bibitem[Kittur et~al\mbox{.}(2011)]%
        {kittur2011crowdforge}
\bibfield{author}{\bibinfo{person}{Aniket Kittur}, \bibinfo{person}{Boris Smus}, \bibinfo{person}{Susheel Khamkar}, {and} \bibinfo{person}{Robert~E Kraut}.} \bibinfo{year}{2011}\natexlab{}.
\newblock \showarticletitle{Crowdforge: Crowdsourcing complex work}. In \bibinfo{booktitle}{\emph{Proceedings of the 24th annual ACM symposium on User interface software and technology}}. ACM, \bibinfo{pages}{43--52}.
\newblock


\bibitem[Kreminski and Martens(2022)]%
        {kreminski-martens-2022-unmet}
\bibfield{author}{\bibinfo{person}{Max Kreminski} {and} \bibinfo{person}{Chris Martens}.} \bibinfo{year}{2022}\natexlab{}.
\newblock \showarticletitle{Unmet Creativity Support Needs in Computationally Supported Creative Writing}. In \bibinfo{booktitle}{\emph{Proceedings of the First Workshop on Intelligent and Interactive Writing Assistants (In2Writing 2022)}}, \bibfield{editor}{\bibinfo{person}{Ting-Hao~'Kenneth' Huang}, \bibinfo{person}{Vipul Raheja}, \bibinfo{person}{Dongyeop Kang}, \bibinfo{person}{John Joon~Young Chung}, \bibinfo{person}{Daniel Gissin}, \bibinfo{person}{Mina Lee}, {and} \bibinfo{person}{Katy~Ilonka Gero}} (Eds.). \bibinfo{publisher}{Association for Computational Linguistics}, \bibinfo{address}{Dublin, Ireland}, \bibinfo{pages}{74--82}.
\newblock
\urldef\tempurl%
\url{https://doi.org/10.18653/v1/2022.in2writing-1.11}
\showDOI{\tempurl}


\bibitem[Lee et~al\mbox{.}(2024)]%
        {lee2024design}
\bibfield{author}{\bibinfo{person}{Mina Lee}, \bibinfo{person}{Katy~Ilonka Gero}, \bibinfo{person}{John Joon~Young Chung}, \bibinfo{person}{Simon~Buckingham Shum}, \bibinfo{person}{Vipul Raheja}, \bibinfo{person}{Hua Shen}, \bibinfo{person}{Subhashini Venugopalan}, \bibinfo{person}{Thiemo Wambsganss}, \bibinfo{person}{David Zhou}, \bibinfo{person}{Emad~A Alghamdi}, {et~al\mbox{.}}} \bibinfo{year}{2024}\natexlab{}.
\newblock \showarticletitle{A Design Space for Intelligent and Interactive Writing Assistants}. In \bibinfo{booktitle}{\emph{Proceedings of the CHI Conference on Human Factors in Computing Systems}}. \bibinfo{pages}{1--35}.
\newblock


\bibitem[luozhouyan(2024)]%
        {python-string-similarity}
\bibfield{author}{\bibinfo{person}{luozhouyan}.} \bibinfo{year}{2024}\natexlab{}.
\newblock \bibinfo{title}{python-string-similarity}.
\newblock
\newblock
\urldef\tempurl%
\url{{https://github.com/luozhouyang/python-string-similarity#normalized-levenshtein}}
\showURL{%
\tempurl}


\bibitem[Mirowski et~al\mbox{.}(2024)]%
        {mirowski2024robot}
\bibfield{author}{\bibinfo{person}{Piotr Mirowski}, \bibinfo{person}{Juliette Love}, \bibinfo{person}{Kory Mathewson}, {and} \bibinfo{person}{Shakir Mohamed}.} \bibinfo{year}{2024}\natexlab{}.
\newblock \showarticletitle{A Robot Walks into a Bar: Can Language Models Serve as Creativity SupportTools for Comedy? An Evaluation of LLMs’ Humour Alignment with Comedians}. In \bibinfo{booktitle}{\emph{The 2024 ACM Conference on Fairness, Accountability, and Transparency}}. \bibinfo{pages}{1622--1636}.
\newblock


\bibitem[Nighojkar and Licato(2021)]%
        {nighojkar-licato-2021-improving}
\bibfield{author}{\bibinfo{person}{Animesh Nighojkar} {and} \bibinfo{person}{John Licato}.} \bibinfo{year}{2021}\natexlab{}.
\newblock \showarticletitle{Improving Paraphrase Detection with the Adversarial Paraphrasing Task}. In \bibinfo{booktitle}{\emph{Proceedings of the 59th Annual Meeting of the Association for Computational Linguistics and the 11th International Joint Conference on Natural Language Processing (Volume 1: Long Papers)}}. \bibinfo{publisher}{Association for Computational Linguistics}, \bibinfo{address}{Online}, \bibinfo{pages}{7106--7116}.
\newblock
\urldef\tempurl%
\url{https://doi.org/10.18653/v1/2021.acl-long.552}
\showDOI{\tempurl}


\bibitem[Orwig et~al\mbox{.}(2024)]%
        {orwig2024language}
\bibfield{author}{\bibinfo{person}{William Orwig}, \bibinfo{person}{Emma~R Edenbaum}, \bibinfo{person}{Joshua~D Greene}, {and} \bibinfo{person}{Daniel~L Schacter}.} \bibinfo{year}{2024}\natexlab{}.
\newblock \showarticletitle{The language of creativity: Evidence from humans and large language models}.
\newblock \bibinfo{journal}{\emph{The Journal of creative behavior}} \bibinfo{volume}{58}, \bibinfo{number}{1} (\bibinfo{year}{2024}), \bibinfo{pages}{128--136}.
\newblock


\bibitem[Quinn and Bederson(2011)]%
        {quinn2011human}
\bibfield{author}{\bibinfo{person}{Alexander~J Quinn} {and} \bibinfo{person}{Benjamin~B Bederson}.} \bibinfo{year}{2011}\natexlab{}.
\newblock \showarticletitle{Human computation: a survey and taxonomy of a growing field}. In \bibinfo{booktitle}{\emph{Proceedings of the SIGCHI conference on human factors in computing systems}}. \bibinfo{pages}{1403--1412}.
\newblock


\bibitem[Reid(1984)]%
        {reid1984radical}
\bibfield{author}{\bibinfo{person}{Joy Reid}.} \bibinfo{year}{1984}\natexlab{}.
\newblock \showarticletitle{The radical outliner and the radical brainstormer: A perspective on composing processes}.
\newblock \bibinfo{journal}{\emph{TESOL quarterly}} \bibinfo{volume}{18}, \bibinfo{number}{3} (\bibinfo{year}{1984}), \bibinfo{pages}{529--534}.
\newblock


\bibitem[Reimers and Gurevych(2019)]%
        {reimers-2019-sentence-bert}
\bibfield{author}{\bibinfo{person}{Nils Reimers} {and} \bibinfo{person}{Iryna Gurevych}.} \bibinfo{year}{2019}\natexlab{}.
\newblock \showarticletitle{Sentence-BERT: Sentence Embeddings using Siamese BERT-Networks}. In \bibinfo{booktitle}{\emph{Proceedings of the 2019 Conference on Empirical Methods in Natural Language Processing}}. \bibinfo{publisher}{Association for Computational Linguistics}.
\newblock
\urldef\tempurl%
\url{https://arxiv.org/abs/1908.10084}
\showURL{%
\tempurl}


\bibitem[Robinson(1951)]%
        {robinson1951logical}
\bibfield{author}{\bibinfo{person}{William~S Robinson}.} \bibinfo{year}{1951}\natexlab{}.
\newblock \showarticletitle{The logical structure of analytic induction}.
\newblock \bibinfo{journal}{\emph{American Sociological Review}} \bibinfo{volume}{16}, \bibinfo{number}{6} (\bibinfo{year}{1951}), \bibinfo{pages}{812--818}.
\newblock


\bibitem[Roemmele(2021)]%
        {roemmele2021inspiration}
\bibfield{author}{\bibinfo{person}{Melissa Roemmele}.} \bibinfo{year}{2021}\natexlab{}.
\newblock \showarticletitle{Inspiration through Observation: Demonstrating the Influence of Automatically Generated Text on Creative Writing}.
\newblock \bibinfo{journal}{\emph{arXiv preprint arXiv:2107.04007}} (\bibinfo{year}{2021}).
\newblock


\bibitem[slab(2024)]%
        {quill-delta}
\bibfield{author}{\bibinfo{person}{slab}.} \bibinfo{year}{2024}\natexlab{}.
\newblock \bibinfo{title}{Delta}.
\newblock
\newblock
\urldef\tempurl%
\url{{https://github.com/slab/delta}}
\showURL{%
\tempurl}


\bibitem[Swales(2017)]%
        {swales2017concept}
\bibfield{author}{\bibinfo{person}{John~M Swales}.} \bibinfo{year}{2017}\natexlab{}.
\newblock \showarticletitle{The concept of discourse community: Some recent personal history}. In \bibinfo{booktitle}{\emph{Composition Forum}}, Vol.~\bibinfo{volume}{37}.
\newblock


\bibitem[Tian et~al\mbox{.}(2024)]%
        {tian2024large}
\bibfield{author}{\bibinfo{person}{Yufei Tian}, \bibinfo{person}{Tenghao Huang}, \bibinfo{person}{Miri Liu}, \bibinfo{person}{Derek Jiang}, \bibinfo{person}{Alexander Spangher}, \bibinfo{person}{Muhao Chen}, \bibinfo{person}{Jonathan May}, {and} \bibinfo{person}{Nanyun Peng}.} \bibinfo{year}{2024}\natexlab{}.
\newblock \showarticletitle{Are Large Language Models Capable of Generating Human-Level Narratives?}
\newblock \bibinfo{journal}{\emph{arXiv preprint arXiv:2407.13248}} (\bibinfo{year}{2024}).
\newblock


\bibitem[Wenger(2000)]%
        {Wenger2000}
\bibfield{author}{\bibinfo{person}{Etienne Wenger}.} \bibinfo{year}{2000}\natexlab{}.
\newblock \showarticletitle{Communities of Practice and Social Learning Systems}.
\newblock \bibinfo{journal}{\emph{Organization}} \bibinfo{volume}{7}, \bibinfo{number}{2} (\bibinfo{year}{2000}), \bibinfo{pages}{225--246}.
\newblock
\urldef\tempurl%
\url{https://doi.org/10.1177/135050840072002}
\showDOI{\tempurl}
\showeprint{https://doi.org/10.1177/135050840072002}


\bibitem[Yujian and Bo(2007)]%
        {4160958}
\bibfield{author}{\bibinfo{person}{Li Yujian} {and} \bibinfo{person}{Liu Bo}.} \bibinfo{year}{2007}\natexlab{}.
\newblock \showarticletitle{A Normalized Levenshtein Distance Metric}.
\newblock \bibinfo{journal}{\emph{IEEE Transactions on Pattern Analysis and Machine Intelligence}} \bibinfo{volume}{29}, \bibinfo{number}{6} (\bibinfo{year}{2007}), \bibinfo{pages}{1091--1095}.
\newblock
\urldef\tempurl%
\url{https://doi.org/10.1109/TPAMI.2007.1078}
\showDOI{\tempurl}


\bibitem[Znaniecki(1934)]%
        {znaniecki1934method}
\bibfield{author}{\bibinfo{person}{Florian Znaniecki}.} \bibinfo{year}{1934}\natexlab{}.
\newblock \bibinfo{booktitle}{\emph{The method of sociology}}.
\newblock \bibinfo{publisher}{Farrar \& Rinehart}.
\newblock


\end{thebibliography}

\appendix
\section{Prompts Used in \gptplot\label{app:gpt-plot-promopt}}
We asked GPT-3~\cite{NEURIPS2020_1457c0d6} to produce a follow-up story arc using the prompt:
\begin{quote}
    \textit{Given the previous story: \texttt{\small[story-snippet]}}.
    
    \textit{Follow the instruction: \texttt{\small[instruction]} to describe the follow-up story arc using 50 words.}
\end{quote}
where \texttt{\small[story-snippet]} and \texttt{\small[instruction]} are the placeholder for the selected snippet and the instruction.

\end{document}